\documentclass[10pt, conference, compsocconf]{IEEEtran}
\usepackage{graphicx}
\usepackage{subfig}
\usepackage{amsfonts}
\usepackage{amssymb,amsmath}
\usepackage{euscript}
\usepackage{color}
\usepackage{url}
\usepackage{algorithm}
\usepackage{algpseudocode}
\usepackage[marginpar]{todo} % il comando \todos \`e qui in fondo
\usepackage[justification=centering]{caption}
\usepackage{balance}

\DeclareMathOperator{\hash}{\mathrm{Hash}}
\DeclareMathOperator{\cheat}{\mathrm{cheat}}
\newtheorem{definition}{Definition}
\newcommand{\mr}{Map\-Reduce}

\begin{document}
\title{Distributed CTL Model Checking in the Cloud}
\author{\IEEEauthorblockN{Carlo Bellettini, Matteo Camilli, Lorenzo Capra, Mattia Monga}
\IEEEauthorblockA{Dept. of Computer Science\\
Universit\`a degli Studi di Milano\\
Milano, Italy\\ 
\{bellettini, camilli, capra, monga\}@di.unimi.it}}
\maketitle

\begin{abstract}
  The recent extensive availability of ``big data'' platforms calls
  for a more widespread adoption by the formal verification
  community. In fact, formal verification requires high performance
  data processing software for extracting knowledge from the
  unprecedented amount of data which come from analyzed systems.
  Since cloud based computing resources have became easily accessible,
  there is an opportunity for verification techniques and tools to
  undergo a deep technological transition to exploit the new available
  architectures. This has created an increasing interest in
  parallelizing and distributing verification techniques.  In this
  paper we introduce a distributed approach which exploits techniques
  typically used by the ``big data'' community to enable verification
  of Computation Tree Logic (CTL) formulas on very large state spaces
  using distributed systems and cloud computing facilities. The
  outcome of several tests performed on benchmark specifications are
  presented, thus showing the convenience of the proposed approach.
\end{abstract}

%\begin{IEEEkeywords}
%Model-Checking; CTL; Big Data; Distributed Algorithms; MapReduce; Cloud Computing;
%\end{IEEEkeywords}

\section{Introduction}
\label{sec:intro}
Ensuring the correctness of software and hardware products is an issue of great importance.
%, as failures can even have fatal consequences in safety-critical systems.
This has led to an increased interest in applying formal methods and verification techniques in order to ensure correctness of developed systems. Among the most successful techniques that are widely used in both research and industry is \emph{model checking}.
Model checking of dynamic, concurrent and real-time systems has been
the focus of several decades of software engineering research. One of
the most challenging task in this context is the development of tools
able to cope with the complexity of the models needed in the analysis
of real word examples. In fact, the main obstacle that model checking
faces is the state explosion problem \cite{Valmari98}: The number of
global states of a concurrent system with multiple processes can be
enormous. It increases exponentially in both the number of processes
and the number of components per process. The most significant
contributions the research has provided in order to cope with this
problem are symbolic model checking with ordered binary decision
diagrams \cite{Burch92}, partial order reduction techniques
\cite{Alur97}, and bounded model checking \cite{Latvala04}. 

These breakthrough techniques have enabled the analysis of systems with a fairly big
number states. Nevertheless, taking advantage of a distributed
environment is still important to cope with real world problems. The
idea is to increase the computational power and a larger available
memory, by using a cluster of computers. The use of networks of
computers can provide the resources required to achieve verification
of models representing real world examples. Unfortunately, this last
approach requires several skills which---while common in the ``big
data'' community---are still rather rare in the ``formal methods''
community.

In fact, our recent works were focused on the connection between formal
methods in software engineering and big data approaches
\cite{Camilli13, Camilli12, Camilli12-2}. The analysis of very complex
systems certainly falls in this context, although formal verification
has so far poorly explored by big data scientists. We believe,
however, the challenges to be tackled in formal verification can
benefit a lot from results and tools available for big data access and
management. In fact formal verification requires several different
skills: On the one hand, one needs an adequate background on formal
methods in order to understand specific formalisms and proper
abstraction techniques for modeling and interpreting the analysis
results; On the other hand, one should also strive to deploy this
techniques into software tools able to analyze large amount of data
very reliably and efficiently similarly to ``big data''
projects. Recent approaches have shown the convenience of employing
distributed memory and computation to manage large amount of reachable
states, but unfortunately exploiting these results requires further skills in developing complex applications with knotty communication and synchronization issues. In particular, adapting an application for exploiting the scalability provided by cloud computing facilities as the Amazon
Cloud Computing platform \cite{amazonElasticMapReduce} might be a
daunting task without the proper knowledge of the subtleties of
data-intensive and distributed analyses.

In this paper, we try to further reduce the gap between these
different areas of expertise by providing a distributed CTL
(\emph{Computation Tree Logic}) model checker, which exploits
computational models typically used to tackle big data problems. Our
software tool is built on top of \textsc{Hadoop \mr}
\cite{Dean08,hadoop} and can be easily specialized to deal with the
verification of CTL formulas on very large state spaces coming from
different kinds of formalisms (e.g., different kinds of Petri Nets,
Process Algebra etc.), thus it is suitable for simplifying the task of
dealing with a large amount of reachable states by exploiting large
clusters of machines.  The \mr\ programming model, which has become
the \textit{de facto} standard for large scale data-intensive
applications, has provided researchers with a powerful tool for
tackling big-data problems in different areas
\cite{Lin10,Camilli13,Cheng06,Biswanath09}. We firmly believe that
explicit state model checking could benefit from a distributed \mr\ based
approach, but the topic has not been yet explored as far as we
know. Exposing this issue to scientists with different backgrounds
could stimulate the development of new interesting and more efficient
solutions.
%The rest of the paper is organized as
%follows. Section~\ref{sec:ctl} gives a brief background on
%CTL; Section~\ref{sec:fixedpoint} describes an approach to model check CTL formulas;
%Section~\ref{sec:disalgo} describes our distributed approach;
%Section~\ref{sec:experiments} describes some experimental results;
%Section~\ref{sec:rel} gives an overview on related works; finally
%Section~\ref{sec:conc} reports our conclusions.

\section{Computation Tree Logic}
\label{sec:ctl}
CTL \cite{Clarke81} is a branching-time logic which models time as a tree-like structure where each moment can be followed by several different possible futures. In CTL each basic temporal operator (\textit{i.e.,} either $X$, $F$, $G$) must be immediately preceded by a path quantifier (\textit{i.e.,} either $A$ or $E$). In particular, CTL formulas are inductively defined as follows: 
$$\phi ::= p \ | \ \neg \phi \ | \ \phi \vee \phi \ | \ A\psi \ | \ E\psi \ (state \ formulas)$$
$$\psi ::= X\phi \ | \ F\phi \ | \ G\phi \ | \ \phi U \phi \ (path \ formulas)$$

Where $p \in AP$, the set of atomic propositions.
The universal path operator $A$ and the existential path operator $E$
express respectively that a property is valid for all paths and for some paths. The temporal operators next $X$ and until $U$ express respectively that a property is valid in the next state, and that a property is valid until another property becomes valid.
The interpretation of a CTL formula is defined over a \emph{Kripke
  structure} (\textit{i.e,} a \emph{state transition system}).
A Kripke structure is made up by a finite set of states, a set of transitions (\textit{i.e.,} a relation over the states), and a labeling function which assigns to each state the set of atomic propositions that are true in this state. Such a model describes the system at any point in time represented by states; the transition relation describes how the system evolves from a state to another over one time step. The formal definition is the following.
\begin{definition}[Kripke structure]
A Kripke structure $T$ is a quadruple $\langle S, S_0, R, L\rangle$, where:
\begin{enumerate}
\item $S$ is a finite set of states.
\item $S_0$ is the set of initial states.
\item $R \subseteq S \times S$ is a a total transition relation, that
  is: $\forall s \in S \ \exists s' \in S \ \text{such that} \ (s,s') \in R$
\item $L : S \rightarrow 2^{AP}$ labels each state with the set of atomic propositions that hold in that state.
\end{enumerate}
\end{definition}
Note that the third point imposes the \emph{seriality} of the
transition relation. This means that the system cannot have deadlock
states. This condition can be always achieved easily by adding into the system a state of ``error'' (with one outgoing transition directed to itself) from which the system cannot get out once reached.

A path $\sigma$ in $T$ from a state $s_0$ is an infinite sequence of states $\sigma = s_0s_1s_2\dots$ where $\forall i \geq 0, \ (s_i,s_{i+1}) \in R$.

\begin{definition}[Satisfiability]
Given a CTL formula $\phi$ and a state transition system $T$ with $s
\in S$, we say that $T$ satisfy $\phi$ in the state $s$ (written as $T \models_s \phi$) if:
\begin{itemize}
\item $T \models_s p$ iff $p \in L(s)$.
\item $T \models_s \neg \phi$ iff $T \not\models_s \phi$.
\item $T \models_s \phi \wedge \psi$ iff $(T \models_s \phi \wedge T \models_s \psi)$.
\item $T \models_s \phi \vee \psi$ iff $(T \models_s \phi \vee T \models_s \psi)$.
\item $T \models_s EX\phi$ iff $\exists t$ such that $R(s, t) \wedge T \models_t \phi$.
\item $T \models_s EG\phi$ iff $\exists$ a path $s_0s_1s_2\dots$ such
  that:\\ $\forall i\geq0, T\models_{s_i}\phi$.
\item $T \models_s E[\phi U \psi]$ iff $\exists$ a path $s_0s_1s_2\dots$ such that:\\ $\exists i\geq0, (T \models_{s_i} \psi) \wedge (T \models_{s_j} \phi \ \forall j < i)$.
\end{itemize}
\end{definition}

We can also write $T \models \phi$ which means that $T$ satisfies
$\phi$ in all the initial states of the system.

It can be shown that any CTL formula can be written in terms of $\neg,
\vee, EX , EG$, and $EU$, for example $AX\phi$ is $\neg EX \neg \phi$,
$EF\phi$ is $E[True \ U \ \phi]$, and so forth. The possible
combinations are only eight:

$$AX, EX, AF, EF, AG, EG, AU, EU$$

\begin{figure*}[htbf]
\centering
\includegraphics[width=1.8\columnwidth]{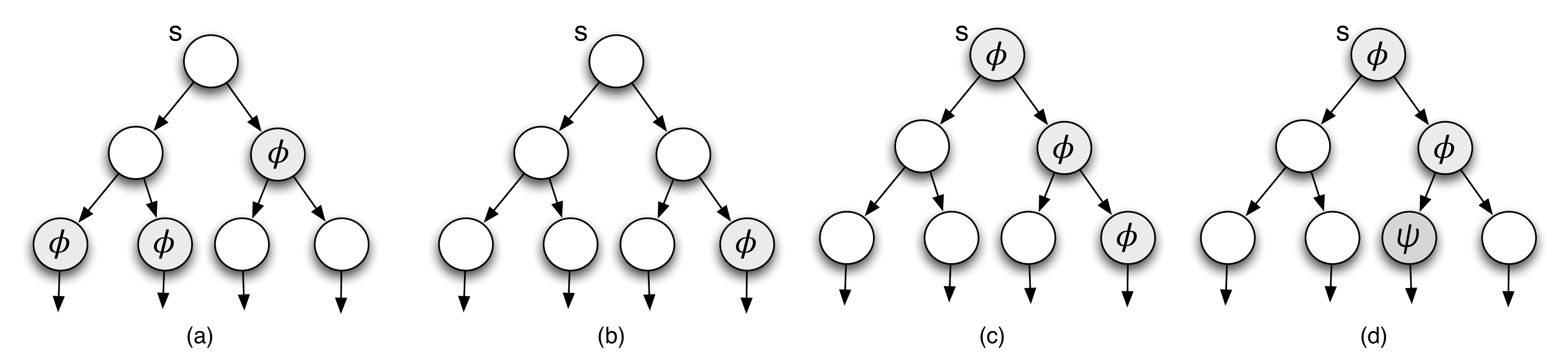}
\caption{(a) $T \models_s AF\phi$; (b) $T \models_s EF\phi$;  (c) $T \models_s EG\phi$;  (d) $T \models_s E[\phi U \psi]$ }

\label{fig:trees}
\end{figure*}

The semantics of some widely used CTL operators is exemplified in Figure~\ref{fig:trees}.

\begin{definition}[Model Checking]
  Let $T$ be a Kripke structure and let $\phi$ be a CTL formula. The
  model checking problem is to find all the states $s \in S$ such that
  $T \models_s \phi$.
\end{definition}

\section{Fixed-Point Algorithms}
\label{sec:fixedpoint}
One of the existing model-checking algorithms is based on fixed-point
characterizations of the basic temporal operators of CTL (similar
ideas can be used for LTL model checking) \cite{Clarke99}. Let $T =
\langle S,S_0,R,L \rangle$ be a Kripke structure. The set $\mathcal{P}(S)$ of all subsets of $S$ forms a lattice under the set inclusion ordering. For convenience, we identify each state formula with the set of states in which it is true. For example, we identify the formula \emph{false} with the empty set of states, and we identify the formula \emph{true} with $S$ (the set of all states). Each element of $\mathcal{P}(S)$ can be viewed both as a set of states and as a state formula (a predicate). Formally, given a CTL formula $\phi$ we can define:
$$[\![ \phi ]\!]_T :=\{s \in S \ : \ T \models_s \phi \}$$
This way, we can associate set operators to boolean connectors:
$$[\![ \phi_1 \wedge \phi_2 ]\!] = [\![ \phi_1 ]\!] \cup [\![ \phi_2 ]\!], \ [\![ \phi_1 \vee \phi_2 ]\!] = [\![ \phi_1 ]\!] \cap [\![ \phi_2 ]\!],$$
$$[\![ \neg \phi ]\!] = S \ \backslash \ [\![ \phi ]\!] $$
The set of states identified by the temporal operator $EX$, can be defined trivially if we consider the counterimage with respect to the relation $R$. Given $W \in \mathcal{P}(S)$:
$$R^-(W) := \{ s \in S \ : \ \exists s' (R(s,s') \wedge s' \in S)\}$$
Thus we can verify easily that the following holds:
$$[\![ EX\phi ]\!]_T = R^-([\![ \phi ]\!]_T)$$
Let's now consider a function $\tau : \mathcal{P}(S) \rightarrow \mathcal{P}(S)$ called \emph{predicate transformer}.
\begin{definition}[Fixed-Point]
We say that a state formula $X$ is the \emph{least fixed-point}
$\mu_X$ (or respectively the \emph{greatest fixed-point} $\nu_X$) of a predicate transformer $\tau$ iff (1) $X = \tau(X)$, and (2) for all state formulas $X'$, if $X' = \tau(X')$, then $X \subseteq X'$ (respectively $X \supseteq X'$).
\end{definition}
\begin{definition}[Monotonic Predicate Transformer]
A predicate transformer $\tau$ is \emph{monotonic} iff for all $X,X' \in \mathcal{P}(S)$ $X \subseteq X'$ implies $\tau(X) \subseteq \tau(X')$.
\end{definition}
A monotonic predicate transformer on $\mathcal{P}(S)$ always has a least fixed-point and a greatest fixed-point (by Tarski's Fixed-Point Theorem \cite{Tarski55}). The temporal operators $EG$ and $EU$ can each be characterized respectively as the greatest and the least fixed-point of two different monotonic predicate transformers:
\begin{equation} \label{eq:pt1}
[\![ EG\phi ]\!]_T = \nu_X([\![ \phi ]\!]_T \cap R^-(X))
\end{equation}
\begin{equation} \label{eq:pt2}
[\![ E[\phi U \psi] ]\!]_T = \mu_X([\![ \psi ]\!]_T \cup ([\![ \phi ]\!]_T \cap R^-(X)))
\end{equation}

We can calculate the least fixed-point of a monotonic predicate transformer: $\mu_X(\tau(X))$ as follows. We define $X_0 = \emptyset$ and $X_i = \tau(X_{i+1})$ for $i \geq 1$. We first compute $X_1$, then $X_2$, then $X_3$, and so forth, until we find a $k$ such that $X_k = X_{k-1}$. It can be proved that the $X_k$ computed in this manner is the least fixed-point of $\tau$. To compute the greatest fixed-point, we follow a similar procedure but starting from $S$. Pseudocode for this procedure is shown by Algorithm \ref{alg:lfp}.
\begin{algorithm}[h!]
\caption{Least Fixed-Point Procedure}
\label{alg:lfp}
\begin{algorithmic}[1]
%\Require $\forall r',r'' \in R \cup \{r\},~ r' \perp r''$
%\Ensure $r ~valid ~direct ~co-location$
\Function{Lfp}{$\tau$}
\State $X := \emptyset$
\While{$X \neq \tau(X)$}
\State $X := \tau(X)$
\EndWhile
\State \Return $X$
\EndFunction
\end{algorithmic}
\end{algorithm}

\section{Distributed Model Checking Algorithms}
\label{sec:disalgo}

We now recall briefly the \mr\ computational model (the basis on top
of which our application is built) and later on we present our
distributed approach in we used the fixed-point algorithms to exploit
distributed and ``cloud'' facilities. The distributed algorithms
presented in this section aim just at computing formulas of type $EX$,
$EG$, and $EU$ because any CTL formula can be reformulated in terms of
these three basic operators (see \ref{sec:ctl}).

\subsection{\mr}
\label{sec:mapred}
\begin{figure}
\centering
\includegraphics[width=0.9\columnwidth]{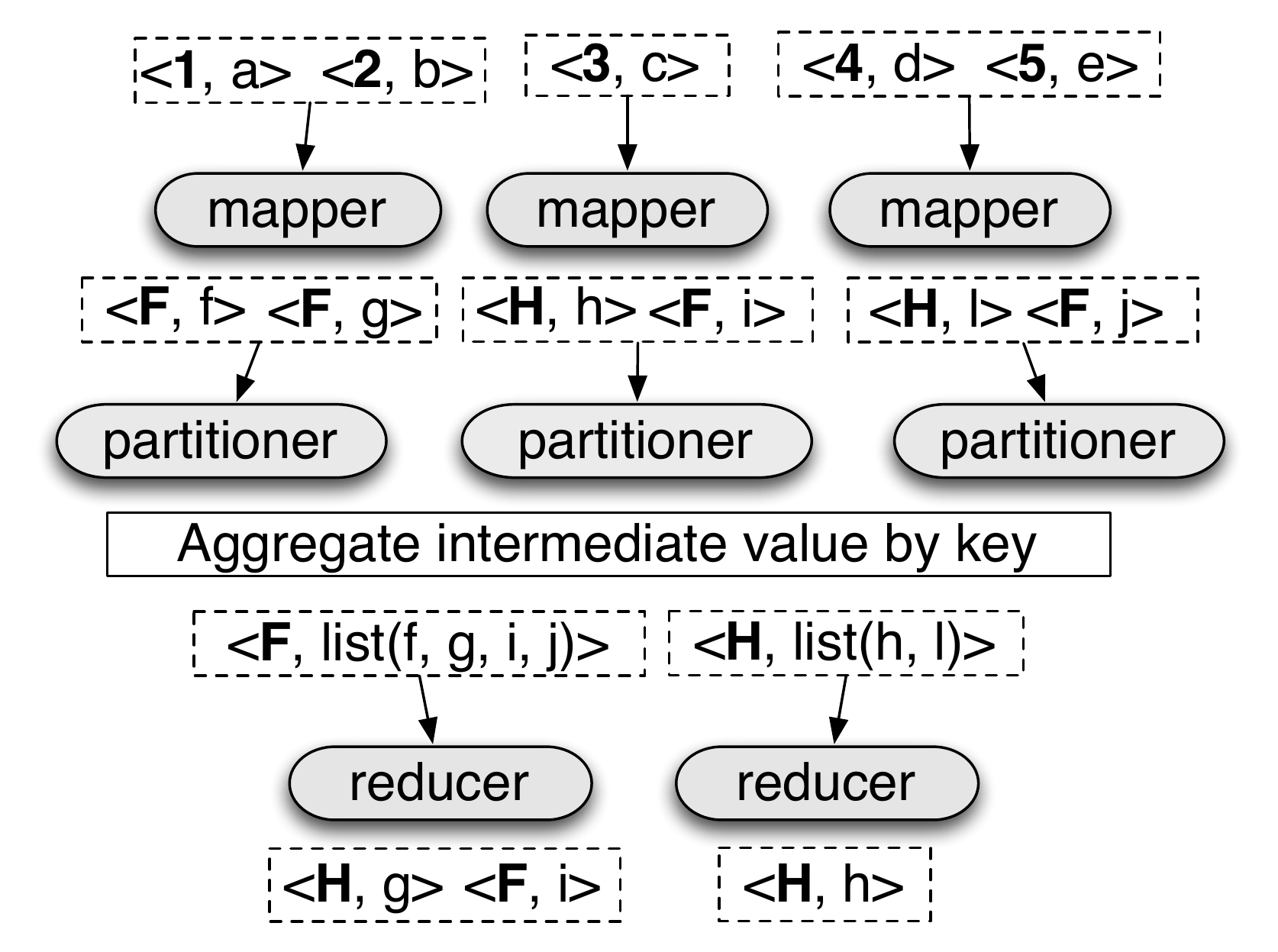}
\caption{The \mr{} model: the keys are in \textbf{bold}.}
\label{fig:mapreduce}
\end{figure}

\mr{} relies on the observation that many information processing
activities have the same basic design: a same operation is applied
over a large number of records (\textit{e.g.,} database records, or
vertices of a graph) to generate partial results, which are then
aggregated to compute the final output.
The \mr\ model consists of two functions: The ``map'' function turns
each input element into zero or more key-value pairs.
A ``key'' is not unique, in fact many pairs with a given key could be generated
from the Map function;
The ``reduce'' function is applied, for each key, to its associated list of values.
The result is a key-value pair consisting of whatever is produced
by the Reduce function applied to the list of values.
Between these two main phases the system sorts the key-value pairs by key,
and groups together values with the same key.
This two-step processing structure is presented in Figure~\ref{fig:mapreduce}.

%In the \mr\ model users create
%their own applications through a ``map'' function which specifies
%per-record computations, and a ``reduce'' function which specifies the
%aggregation of map computations: both operate in parallel on
%key-value pairs which represent the input of the problem. The mapper
%is applied to every input key-value pair to generate an arbitrary
%number of intermediate key-value pairs. The reducer is then applied to
%all values associated with the same intermediate key to generate an
%arbitrary number of final key-value pairs as output.

%Under the \mr{} programming model, a developer needs to provide implementations of the mapper and reducer. On top of a distributed file system \cite{Ghemawat03},
The execution framework handles transparently all
non-functional aspects of execution on big clusters. It is responsible, among other things, for scheduling (moving code to data), handling faults, and the large distributed sorting and shuffling needed between the map and reduce phases since intermediate key-value pairs must be grouped by key.
%As an optimization, \mr{} supports the use of ``combiners'', which are similar to reducers except that they operate on the output of single mappers. Combiners operate in isolation on each node in the cluster after a mapper and cannot use partial results from other nodes.
%They allow a programmer to aggregate partial results (i.e., intermediate key-value pairs), thus reducing network traffic. In cases where an operation is both associative and commutative, reducers can directly serve as combiners, although in general they are not interchangeable.
The ``partitioner'' is responsible for dividing up the intermediate
key space and assigning intermediate key-value pairs to reducers. The
default partitioner computes a hash function on the value of the key modulo the number of reducers.
%This does not guarantee good load balance because the distribution of values associated with the same key may be highly skewed.

\subsection{Distributed State Space Generation}
\label{sec:x}
This task builds the reachability graph $T$ of a given model in a distributed fashion. The idea underlying a distributed algorithm for state space exploration is to use multiple computational units to perform the exploration of different parts of the whole state space in parallel. The task is typically performed by using classical parallel \emph{Worker}s algorithms \cite{Kristensen04}: States are partitioned among workers by means of a static hash function. The workers explore successor states and assign them to the proper computational units. Communication among different machines is implemented through message passing. Since the partitioning of the state space is a critical issue, different load balancing techniques and compact states representation \cite{Nicol97,Kumar05,Garavel01} were studied.
Recent approaches has shown also the convenience of exploiting big
data approaches and cloud computing facilities in order to accomplish
this task. In particular the \emph{MaRDiGraS} \cite{Camilli13}
framework could be employed to implement distributed state space
builders for different formalisms. Given a cluster size of $n$
machines, a \emph{MaRDiGraS} based application generates $n$ files
$F_1, F_2, ..., F_n$ containing the whole state space, partitioned
into $n$ different sets. The set of states emitted by the $i$\/th
computational unit is $S_i=\{s \in S : \hash(f(s)) = i\}$, where $S$
is the set of reachable states, $f$ is a user supplied function and
$f(s)$ computes specific features on states such that the equality of
the evaluation of these features is a necessary condition for having
an inclusion/equality relationship among states. What makes this
representation interesting and suitable for further analysis by using
our distributed approach is in the transitions management (the $R$
relation). In particular each state stores locally all incoming
transitions as a list of state identifiers, therefore, given a set of state $W$, $R^-(W)$ can be easily computed:
$$\bigcup_{i=1}^{n}s_i \in S : (id(s_i) \in R^-(s_j), \forall s_j \in W)$$

It is worth noting that the set of predecessor states' identifiers should be immediately available inside state definition because our \mr\ based approach exploits the evaluation of $R^-$ as a basic operation without communication among computational units.

In order to apply our distributed fixed-point algorithms the analyzed transition system must preserve the \emph{seriality} of the transition relation (introduced in section \ref{sec:ctl}). If not, the \emph{MaRDiGraS} framework can add an output file containing a single ``error'' state where the list of incoming transitions is made up by itself and all deadlock states.

\subsection{$EX$ Formulas}
\label{sec:ex}
To compute $[\![ EX\phi ]\!]_T$, we assume that the set of states satisfying $\phi$ is already computed. Thus $\phi$ can be either a formula locally evaluable or a more complex sub-formula evaluated previously. We can deploy this operation into a single \mr\ job where the predecessor states of the $[\![ \phi ]\!]_T$ set are evaluated in parallel. The input of this distributed computation is two different sets of files. 
The first set contains all states belonging to $S \setminus [\![ \phi
]\!]_T$, the second contains all states belonging to $[\![ \phi
]\!]_T$. This way all the mappers can evaluate  and emit in parallel
the identifiers of the states belonging to $R^-([\![ \phi ]\!]_T)$. As
shown by Algorithm~\ref{alg:ex}, the \texttt{Map} function emits the
identifiers of these states associated with an empty value
$\perp$. Then the shuffle phase groups together  all the values with the same identifier, so that the \texttt{Reduce} function can emit the final result by checking whenever the empty value was passed into the input list.

\begin{algorithm}[h!]
\caption{\mr\ algorithm for evaluating $EX\phi$}
\label{alg:ex}
\begin{algorithmic}[1]
%\Require $\forall r',r'' \in R \cup \{r\},~ r' \perp r''$
%\Ensure $r ~valid ~direct ~co-location$
\Function{Map}{$k, s$}
\If{$s \in [\![ \phi ]\!]_T$}
\For{$e \in R^-(s)$}
\State $emit(e, \perp)$
\EndFor
\EndIf
\State $emit(k, s)$
\EndFunction

\Function{Reduce}{$k, list := [s_1,s_2,...]$}
\If{$\perp \in list$}
\State $s := s' \in list \ s.t. \ s' \neq \perp$
\State $emit(k, s)$
\EndIf
\EndFunction
\end{algorithmic}
\end{algorithm}

\subsection{$EG$ Formulas}
\label{sec:eg}
As for the previous formula, to compute $[\![ EG\phi ]\!]_T$, we assume that the set of states satisfying $\phi$ is already computed. The evaluation of the final result is a bit more complex than the previous case. Our approach is based on the greatest fixed-point characterization of the monotonic predicate transformer introduced in (\ref{eq:pt1}). Thus we apply an iterative \mr\ algorithm, where at each iteration we compute the predicate transformer on the output of the previous iteration until we reach the fixed-point.
Algorithm~\ref{alg:eg} shows the \texttt{Map} and the \texttt{Reduce} functions employed within the job iterations. The input of each \mr\ job is made up by a set of files containing $[\![ \phi ]\!]_T$ and another set of files $X$ representing the current evaluation of the formula.
Since the first iteration should start from $X=S$ and $R^-(S)=S$, we already know the result of the first evaluation of the predicate transformer introduced in (\ref{eq:pt1}), thus we start directly from the second iteration by posing $X$ to $[\![ \phi ]\!]_T$. As shown by Algorithm~\ref{alg:eg}, the map phase computes in parallel all the predecessor states and the reduce phase verifies and emits in parallel all predecessors belonging to $[\![ \phi ]\!]_T$. The iterations keep going until the number of key-value pairs given in output by two consecutive jobs becomes equal or we reach the empty set. 

\begin{algorithm}[h!]
\caption{\mr\ for evaluating $EG\phi$}
\label{alg:eg}
\begin{algorithmic}[1]
%\Require $\forall r',r'' \in R \cup \{r\},~ r' \perp r''$
%\Ensure $r ~valid ~direct ~co-location$
\Function{Map}{$k, s$}
\If{$s \in X$}
\For{$e \in R^-(s)$}
\State $emit(e, \perp)$
\EndFor
\EndIf
\If{$s \in [\![ \phi ]\!]_T$}
\State $emit(k, s)$
\EndIf
\EndFunction

\Function{Reduce}{$k, list := [s_1,s_2,...]$}
\If{$\perp \in list \ \wedge \ (s \neq \perp \in list)$}
\State $emit(k, s)$
\EndIf
\EndFunction
\end{algorithmic}
\end{algorithm}

\subsection{$EU$ Formulas}
\label{sec:eu}
As for the previous formulas, to compute $[\![ E[\phi U \psi] ]\!]_T$, we assume that the set of states satisfying the two sub-formulas $\phi$ and $\psi$ are already computed. The approach employed to evaluate this formulas is similar to the previous one, in fact our distributed algorithm is based on the least fixed-point characterization of the monotonic predicate transformer introduced in (\ref{eq:pt2}). The iterative map-reduce algorithm, which uses the \textsc{Map} and the \textsc{Reduce} functions presented by the algorithm \ref{alg:eu}, is employed in order to reach the fixed-point.
The input of each iteration is made up by a set of files $X$ containing the current evaluation of the formula and another set of files containing $[\![ \psi ]\!]_T$. Since the first iteration should start from the empty set, but we already know that the predicate transformer (\ref{eq:pt2}) computed on the input $X=\emptyset$ is $[\![ \psi ]\!]_T$, we start directly from the second iteration posing $X$ to $[\![ \psi ]\!]_T$. The map phase emits in parallel all predecessor states of  $X$ set and forwards all states of $[\![ \psi ]\!]_T$ to reducers. The reduce phase emits in parallel all  predecessor states of $[\![ \phi ]\!]_T$ and all states of $[\![ \psi ]\!]_T$.

\begin{algorithm}[h!]
\caption{\mr\ algorithm for evaluating $E[\phi U \psi]$}
\label{alg:eu}
\begin{algorithmic}[1]
%\Require $\forall r',r'' \in R \cup \{r\},~ r' \perp r''$
%\Ensure $r ~valid ~direct ~co-location$
\Function{Map}{$k, s$}
\If{$s \in X$}
\For{$e \in R^-(s)$}
\State $emit(e, \perp)$
\EndFor
\EndIf
\If{$s \in [\![ \phi ]\!]_T \vee s \in [\![ \psi ]\!]_T$}
\State $emit(k, s)$
\EndIf
\EndFunction

\Function{Reduce}{$k, list := [s_1,s_2,...]$}
\State{$s := s' \in list \ s.t. \ s' \neq \perp$}
\If{$(\perp \in list \ \wedge \ s \neq null) \vee (s \in [\![ \psi ]\!]_T)$}
\State $emit(k, s)$
\EndIf
\EndFunction
\end{algorithmic}
\end{algorithm}

$R^-(X_{i-1}) \subseteq R^-(X_i)$, since Algorithm~\ref{alg:eu} computes $R^-(X)$ for each iteration and $X_{i-1} \subseteq X_i$.
For this reason we implemented an optimized version which computes,
for each iteration, just $R^-(X_i \setminus X_{i-1})$.

\section{Experiments}
\label{sec:experiments}

The experiments described in this section were executed using the
Amazon Elastic \mr\ \cite{amazonElasticMapReduce} on the Amazon Web
Service cloud infrastructure. They were supported by an ``AWS in Education Grant award''~\cite{grant}. In particular all runs have been performed on clusters of various sizes made up by \emph{m2.2xlarge} computational units \cite{amazonElasticMapReduce}.

As a proof of concept we generated three different state spaces, sized
with different order of magnitude. Successively we applied our
distributed algorithms in order to verify three different CTL formulas
(of type $EX$, $EG$ and $EU$) for each state space.
Both models and formulas used during the experiments were introduced in \cite{mcc2013web}.
The models are three Petri Net benchmarks and their state space were generated by means of a \emph{MaRDiGraS} based tool.

\subsection{Shared Memory}
\label{sec:mem}
%This model is taken from the GreatSPN benchmarks \cite{68539,DBLP:journals/corr/abs-1209-2382}.
This P/T net models a system composed of 10 processors which compete for
the access to a shared memory by using a unique shared bus. The number of reachable states of this model is $1.831\times10^6$.
%Each processor can access its local memory using a dedicated local bus and the other memories using a unique shared bus. The processor accessing a remote memory have priority on those accessing their own memory. It is assumed that external access request causes preemption of the owner processor eventually accessing its local memory.
Given the function $m:\mathrm{Place} \rightarrow \mathbb{N}$ which computes the number of tokens for a given place, the three properties verified on this model are:
$$EX[A], \ EG[A], \ E[\mathrm{True} \ U \ A]=EF[A]$$
where:
$$A:= m(\mathrm{Active}) \neq m(\mathrm{Memory}) \vee m(\mathrm{Queue})=m(\mathrm{Active})$$
Despite the generated state space is relatively small, the benefit
gained from our distributed approach grows as the number of states
involved in the verification grows (as shown in
Table~\ref{tab:report}): indeed, the verification of the last formula
$E[\mathrm{True} \ U A]$ scales better than the previous two.

\begin{table}
\caption{Shared memory report}
\label{tab:report}
\centering
{ \begin{tabular}{ | c | c | c | c |}
 \hline
  property & $| [\![ property ]\!]_T |$ & $\#$ machines &  time \emph{(s)}\\
  \hline
  $EX[A]$ & $2.135 \times 10^5$ & 1 & 70\\
  \hline
  $EX[A]$ & $2.135 \times 10^5$ & 2 & 67\\
  \hline
  $EX[A]$ & $2.135 \times 10^5$ & 4 & 50\\
  \hline
  $EX[A]$ & $2.135 \times 10^5$ & 8 & 38\\
  \hline
  \hline
  $EG[A]$ & 0 & 1 & 67\\
  \hline
  $EG[A]$ & 0 & 2 & 55\\
  \hline
  $EG[A]$ & 0 & 4 & 58\\
  \hline
  \hline
  $E[\mathrm{True} \ U A]$ & $1.831 \times 10^6$ & 1 & 1898\\
  \hline
  $E[\mathrm{True} \ U A]$ & $1.831 \times 10^6$ & 2 &  1124\\
  \hline
  $E[\mathrm{True} \ U A]$ & $1.831 \times 10^6$ & 4 &   839\\
  \hline
  $E[\mathrm{True} \ U A]$ & $1.831 \times 10^6$ & 8 &  564\\
  \hline
  $E[\mathrm{True} \ U A]$ & $1.831 \times 10^6$ & 16 &  509\\
  \hline
 \end{tabular}}
\end{table}

\subsection{Dekker}
\label{sec:dekker}
This model represents a 1-safe P/T net of a variant of the Dekker's mutual exclusion algorithm \cite{Dijkstra02} for $N=20$
processes.
The state space generated by this model is an order of magnitude higher than the previous example ($1.153 \times 10^7$ reachable states). 
%Each process has three states. From the initial state, the process executes try and raises its own flag, reaching the second state. From there, if at least one of the other process has a high 
%flag, it withdraws its intent and goes back to the former state.
%The critical section can be accessed if all other process' 
%flag is zero. From this latter state, the process can only exit the critical section.
%Mutual exclusion and deadlock-freedom is guaranteed.
The three properties verified on this model are:
$$EX[B], \ EG[B], \ E[C \ U \ D]$$
where:
\begin{equation}
B:= m(p_{1,18}) \neq m(p_{1,13}) \vee m(p_{0,15})=m(p_{3,18}) \notag
\end{equation}
\begin{equation}
C:=m(flag_{1,18}) \neq m(p_{0,4}) \wedge m(p_{0,17})=m(flag_{1,11}) \notag 
\end{equation}
\begin{equation}
D:=m(p_{0,17}) = m(flag_{1,11}) \notag
\end{equation}

In this case, as shown by Table~\ref{tab:report2} and by the graph shown in Figure~\ref{fig:graphs}(b), the benefits deriving from our distributed approach are clearer. In fact, the evaluation of both the three formulas gets substantially faster by increasing the number of computational units.
The graph shown by Figure~\ref{fig:graphs}(b) (and Figure~\ref{fig:graphs}(d) for the next model), plots the function $\cheat$ defined as follow:
\begin{equation}
\cheat(n) = \frac{\text{exec. time of parallel version with $1$ node}}{\text{exec. time of parallel version with $n$ nodes}}
\label{eq:cheat}
\end{equation}

\begin{table}
\caption{Dekker report}
\label{tab:report2}
\centering
{ \begin{tabular}{ | c | c | c | c |}
 \hline
  property & $| [\![ property ]\!]_T |$ & $\#$ machines & time \emph{(s)}\\
  \hline
   $EX[B]$ & $1.153 \times 10^7$ & 1 & 660\\
  \hline
  $EX[B]$ & $1.153 \times 10^7$ & 2 &  532\\
  \hline
  $EX[B]$ & $1.153 \times 10^7$ & 4 &  241\\
  \hline
  $EX[B]$ & $1.153 \times 10^7$ & 8 &  144\\
  \hline
  $EX[B]$ & $1.153 \times 10^7$ & 16 & 120\\
  \hline
  \hline
  $EG[B]$ & $7.405 \times 10^6$ & 1 &  1567\\
  \hline
  $EG[B]$ & $7.405 \times 10^6$ & 2 & 1356\\
  \hline
  $EG[B]$ & $7.405 \times 10^6$ & 4 & 517\\
  \hline
  $EG[B]$ & $7.405 \times 10^6$ & 8 & 391\\
  \hline
  $EG[B]$ & $7.405 \times 10^6$ & 16 & 287\\
  \hline
  \hline
  $E[C \ U \ D]$ & $5.767 \times 10^6$ & 1 &  1357\\
  \hline
  $E[C \ U \ D]$ & $5.767 \times 10^6$ & 2 & 1063\\
  \hline
  $E[C \ U \ D]$ & $5.767 \times 10^6$ & 4 & 585\\
  \hline
  $E[C \ U \ D]$ & $5.767 \times 10^6$ & 8 & 454\\
  \hline
  $E[C \ U \ D]$ & $5.767 \times 10^6$ & 16  &  372\\
  \hline
 \end{tabular}}
\end{table}

\subsection{Simple Load Balancing}
\label{sec:simple-lbs}
This P/T net represents a simple load balancing system composed of 10 clients, 2 servers, and between these, a load balancer process. 
%The role of clients is to send requests to servers, wait for an answer and get it. Each of the two servers waits for requests (i.e., tokens in place \emph{server\_request}). When such a request arrives it processes it and send a reply to the client. The server then notifies the \emph{lb} process so that this one possibly rebalances requests between servers. Once the \emph{lb} process has acknowledged this
%notification, the server can go back to the idle state. 
The reachability graph generated is very large: $4.060 \times 10^8$ states and $3.051 \times 10^9$ arcs for a total size of 120~GB of data. The three properties verified on this model are:
$$EX[H], \ EG[J], \ E[K \ U \ H]$$
where:
\begin{multline}
H := m(\mathrm{server\_processed}) \neq m(\mathrm{server\_notification}) \wedge \notag \\
m(\mathrm{server\_waiting}) = m(\mathrm{server\_idle}) \notag \\[.3cm]
J := m(\mathrm{client\_idle}) \neq m(\mathrm{client\_waiting}) \notag \\[.3cm]
K :=  m(\mathrm{client\_idle}) \neq m(\mathrm{client\_waiting}) \wedge  \notag \\
m(\mathrm{client\_idle}) = m(\mathrm{client\_request}) \notag \\
\end{multline}

\begin{figure*}[htp]
  \centering
  \caption{Model checking time and cheat of Dekker and Simple load balancing models.}
  \subfloat[Dekker model checking time]{\label{figur:1}\includegraphics[width=8.2cm]{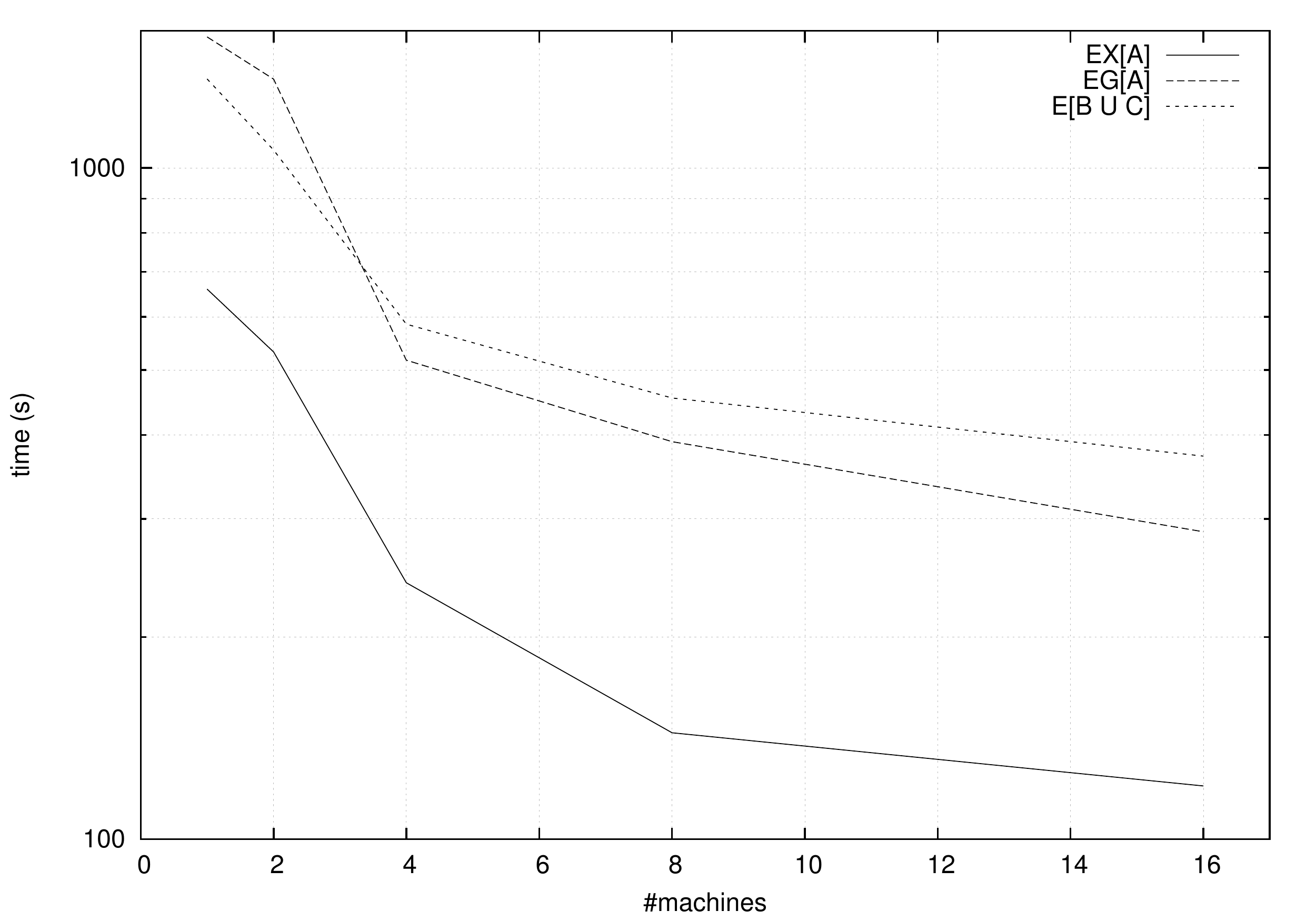}}
  \subfloat[Dekker $\cheat$ (\ref{eq:cheat})]{\label{figur:2}\includegraphics[width=8.2cm]{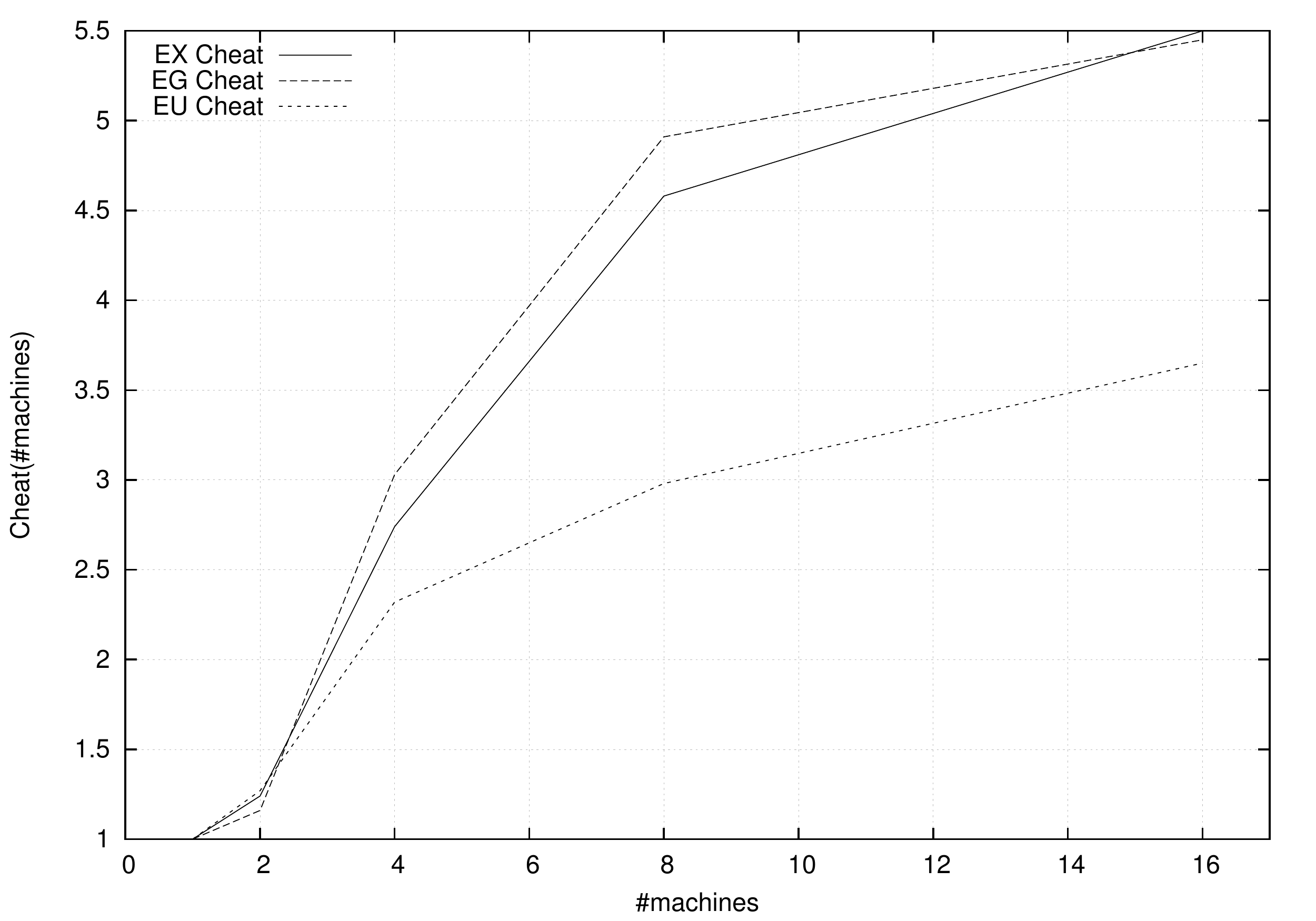}}
  \\
  \subfloat[Simple load balancing model checking time]{\label{figur:1}\includegraphics[width=8.2cm]{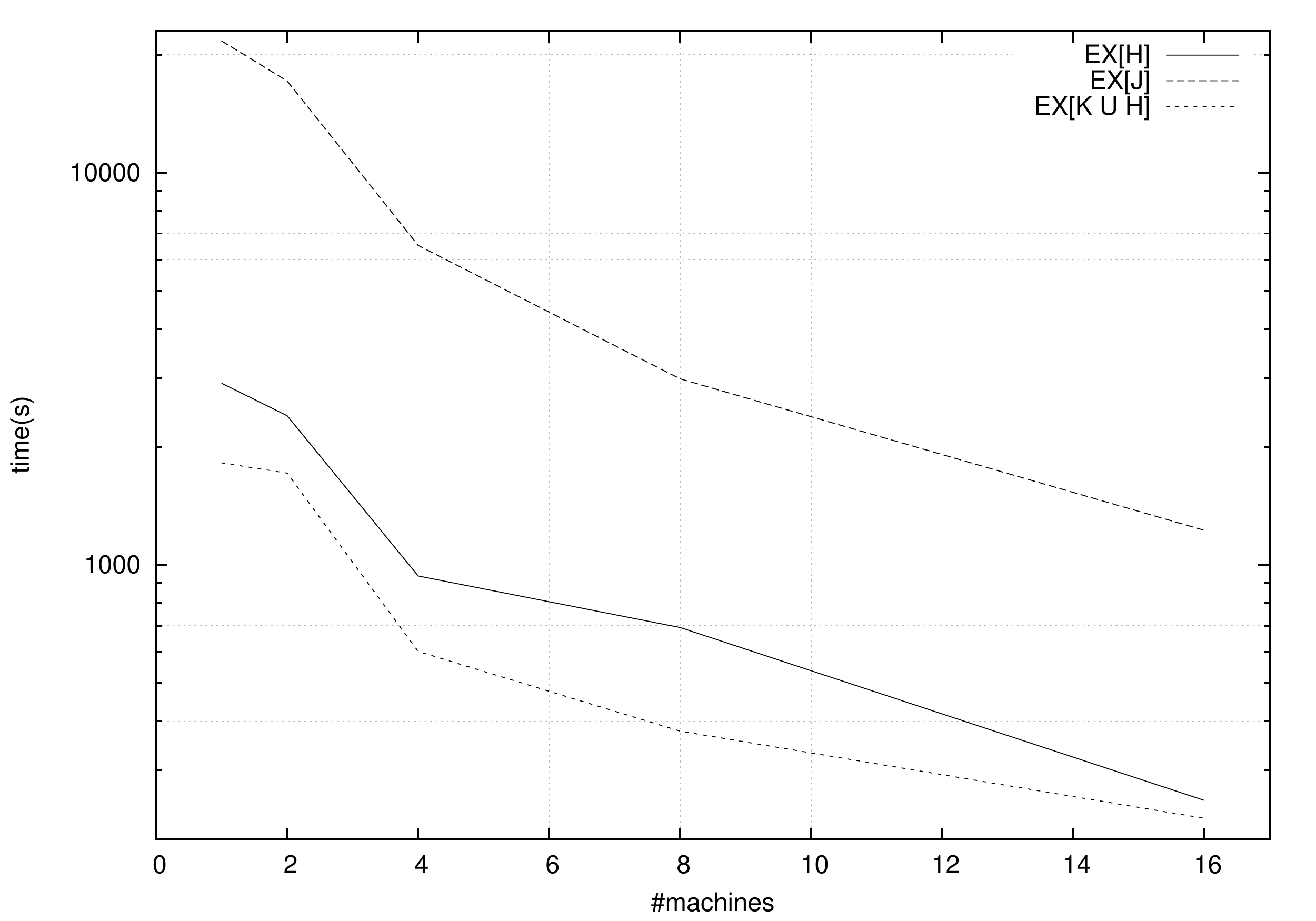}}
  \subfloat[Simple load balancing $\cheat$ (\ref{eq:cheat})]{\label{figur:2}\includegraphics[width=8.2cm]{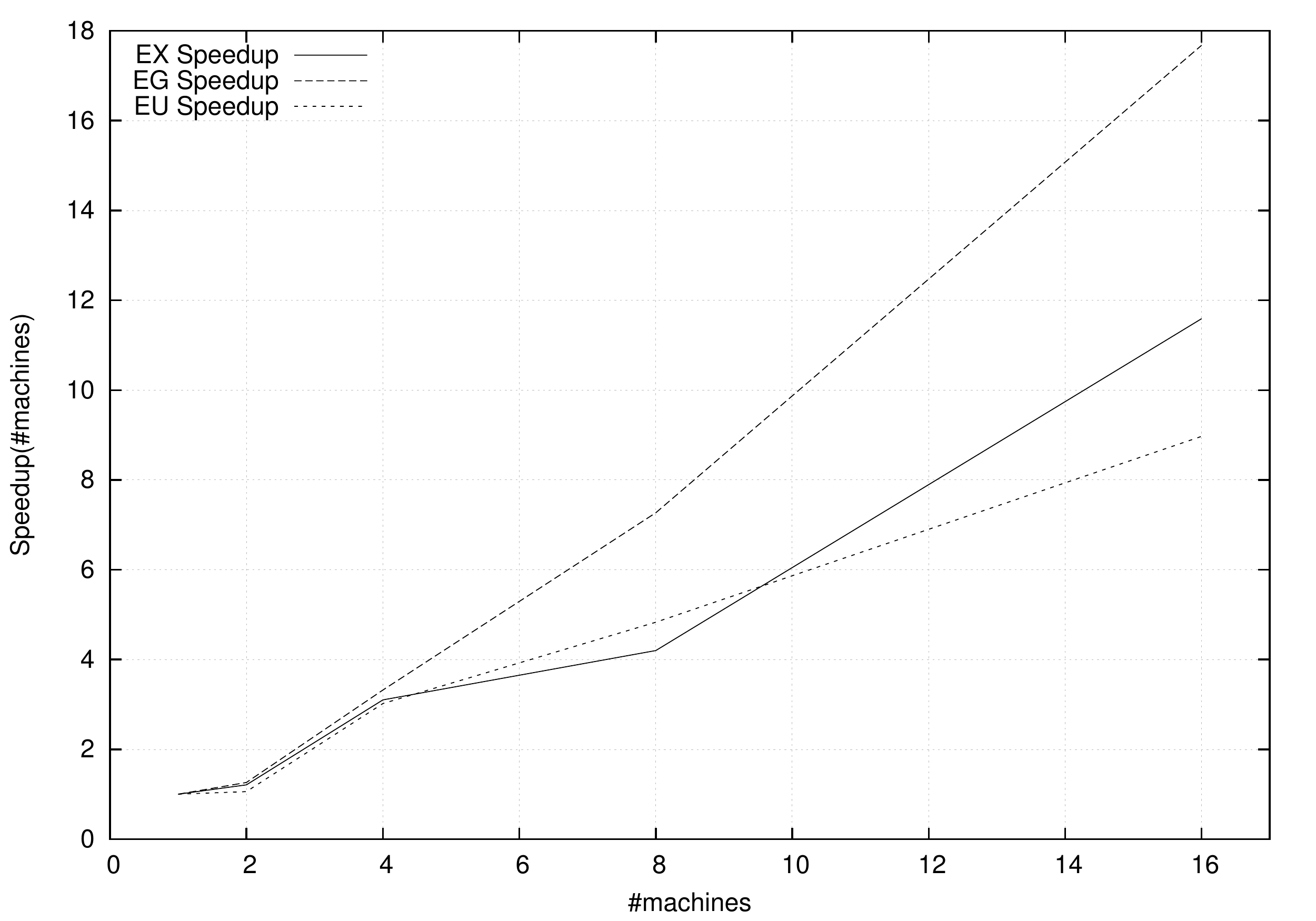}}
\label{fig:graphs}
\end{figure*}

As shown by Table~\ref{tab:report3} and by the graph shown in Figure~\ref{fig:graphs}(d), the benefits deriving from our distributed approach are greater with respect to both previous examples. This points out a clear trend: the major is the complexity of the model to be analyzed, the major is the scalability of our distributed algorithm.
%In fact, the evaluation of both the three formulas gets substantially faster by increasing the number of computational units.
In fact, the $\cheat$ gained during the analysis of this last example
greatly overcome the one gained in the analysis of the Dekker model
(5.5 using 16 machines to evaluate $EX[B]$).  As shown in
Figure~\ref{fig:graphs}(d), in this model we reach a super-linear
speedup during the evaluation of $EG[J]$.

\begin{table}
\caption{Simple load balancing report}
\label{tab:report3}
\centering
{ \begin{tabular}{ | c | c | c | c |}
 \hline
  property & $| [\![ property ]\!]_T |$ & $\#$ machines &  time \emph{(s)}\\
  \hline
  $EX[H]$ & $1.716 \times 10^8$ & 1 & 2908\\
  \hline
  $EX[H]$ & $1.716 \times 10^8$ & 2 & 2401\\
  \hline
  $EX[H]$ & $1.716 \times 10^8$ & 4 &  937\\
  \hline
  $EX[H]$ & $1.716 \times 10^8$ & 8 &  693\\
  \hline
  $EX[H]$ & $1.716 \times 10^8$ & 16 & 251\\
  \hline
  \hline
  $EG[J]$ & $4.060 \times 10^8$ & 1 &  21678\\
  \hline
  $EG[J]$ & $4.060 \times 10^8$ & 2 & 17147\\
  \hline
  $EG[J]$ & $4.060 \times 10^8$ & 4 &  6525\\
  \hline
  $EG[J]$ & $4.060 \times 10^8$ & 8 &  2983\\
  \hline
  $EG[J]$ & $4.060 \times 10^8$ & 16 & 1226\\
  \hline
  \hline
  $E[K \ U \ H]$ & $7.524 \times 10^7$ & 1 & 1821\\
  \hline
  $E[K \ U \ H]$ & $7.524 \times 10^7$ & 2 & 1714\\
  \hline
  $E[K \ U \ H]$ & $7.524 \times 10^7$ & 4 &  602\\
  \hline
  $E[K \ U \ H]$ & $7.524 \times 10^7$ & 8 &  377\\
  \hline
  $E[K \ U \ H]$ & $7.524 \times 10^7$ & 16 & 203\\
  \hline
 \end{tabular}}
\end{table}

\section{Related Work}
\label{sec:rel}
The use of distributed and/or parallel processing to tackle the state explosion problem gained interest in recent years.
In fact, for very complex models, the state space may not completely fit into the main memory
of a single computer and hence model-checking tools becomes very slow or even crash as soon as the memory is exhausted.

\cite{Lerda99,Evangelista11,1240299,Brim04,5698463} discuss
parallel/distributed verification of \emph{Linear Temporal Logic}
(LTL) formulas. They aim at increasing the memory available and
reducing the overall time required by LTL formulas verification by
employing distributed techniques for searching accepting cycles in
B\"ichi automata.  Distributed and parallel model checking of CTL
logic was also proposed. \cite{Brim05} introduced a CTL model checking
technique which works by splitting the given state space into several
``partial state spaces''. Each computer involved in the distributed
computation owns a partial state space and performs a model checking
algorithm on this incomplete structure. To be able to proceed, the
border states are augmented by assumptions about truth values of
formulas and the computers exchange assumptions about relevant states
to compute more precise information.  Other approaches were introduced
in \cite{Bell05,Boukala12}.

The main idea of distributed algorithms for both LTL and CTL model
checking is in fact similar: the state graph is partitioned among the
network nodes, \textit{i.e.,} each network node owns a subset of the
state space.  The differences are in the way the state space is
partitioned (through a \emph{partition function}): this is a crucial
issue. In order to increase performance of the parallel model
checking, it is key to achieve a good load balancing among machines,
meaning that each partition should contain nearly the same number of
states. The performance of these algorithms depends also on the number
of cross-border transitions of the partitioned state space
(\textit{i.e.,} transitions having the source state in a component and
the target state in another component). This number should be as small
as possible, since it has an effect on the number of messages sent
over the network during the analysis \cite{Bourahla05}.  In the
context of LTL model checking, probabilistic techniques to partition
the state space have been used, for example, in
\cite{Lerda99,Stern01}, and a technique that exploits some structural
properties derived from the verified formula has been proposed in
\cite{406260}.

Since our distributed algorithms are quite different from message
passing approaches, the number of cross-border transitions is not a
crucial issue to cope with. The only synchronization point among
computational units is the shuffle phase, where key-value pairs are
sorted and transferred from map outputs to reducers input. Reducing
the number of cross-border transitions may reduce the data exchanged
across the network during this phase. Anyway, this phase is partially
overlapped with the map phase, which means that the shuffling starts
as soon as data become available from mappers without waiting for the
entire map output. Furthermore, since we found experimentally that the
time required by this phase does not dominate the overall time
required by our algorithms, adding a partitioning phase between each
\mr\ iteration could even hurt performances. Nevertheless, we
plan to study further this issue in order to understand better how
partitioning can impact performances of our \mr\ based approach.

Our contribution is a set of parallel algorithms designed for distributed memory architectures and cloud computing platform based on a new emerging distributed paradigm. It is worth noting that departing from the current literature on distributed CTL model checking, we considered an important aspect, sometimes understated: we wanted to completely remove the costs of deploying our application into an end-to-end solution, for this reason we developed our software on top of the consolidated \textsc{Hadoop \mr{}} framework. As far as we now, the effectiveness of a \mr\ based approach, typically employed to solve big data problems, has been not explored so far by the formal verification community. Thus with our work we aim at further reducing the gap between these two different but related areas of expertise.

\section{Conclusion and Future Work}
\label{sec:conc}

In this paper we presented a software framework to model check very
complex systems by applying iterative \mr\ algorithms based on
fixed-point characterizations of the basic temporal operators of CTL. 

Our distributed application exploits techniques typically used by the big data
community and so far poorly explored for this kind of
problem. Therefore we remark a clear connection between formal
verification problems and big data problems conveyed by the recent
widespread accessibility of powerful computing resources.
Despite model checking software tools are so called ``push-button'',
the setup phase required by a distributed application, is far from
being considered such, especially whenever one wants to exploits general purpose ``cloud''
computing facilities. Our framework aims at re-enabling a ``push-button'' mode into
the distributed verification context even when these (complex on themselves) computing
resources are involved. 

Our experiments report that our approach can be used effectively to
analyze state spaces of different orders of magnitude. In particular,
the major is the complexity of the model to be analyzed, the major is
the scalability of our distributed algorithms. In some cases we have
shown a potential for a super-linear speedup.  We believe that this
work could be a further step towards a synergy between two very
different, but related communities: the ``formal methods'' community
and the ``big data'' community. Exposing this issue to scientists with
different backgrounds could stimulate the development of new
interesting and more efficient solutions.
%Concerning future work, we are now employing similar techniques to verify the correctness of real-time systems. The introduction of time metrics and abstraction techniques in the state space representation can indeed complicate this task.

%several questions remains open and require further investigation: for example, could a dynamic programming approach help in choosing partitions and/or thresholds? How the proposed computational model can be optimized when the number of new states gets very small? Are there classes of formalisms for which this approach cannot be used? And how can we adapt it to these classes?

\section*{Acknowledgments}
\label{sec:ack}

The authors would like to thank Amazon.com, Inc. for the ``AWS in Education
Grant'' award which made possible the experiments described in this
paper.

\balance

\bibliographystyle{IEEEtran}
\bibliography{biblio}
%\todos

\end{document}